# Magnetic field driven enhanced ferroelectric switching in self-grown ferroelectric-ferromagnetic composite in the BiFeO$_3$-BaTiO$_3$ multiferroic alloy system

Amit Kumar, Bastola Narayan, Rohit P, and Rajeev Ranjan*

Materials Engineering Department, Indian Institute of Science, Bangalore-560012.

Over the years attempts have been made to compensate for the inherent weaknesses in the bulk state of the multiferroic BiFeO$_3$, such as high leakage current and the absence of ferromagnetic correlation, and exploit its magnetoelectric potential by forming solid solutions with other perovskites. Studies in the recent few years have shown that alloying of BiFeO$_3$ with BaTiO$_3$, both with and without additives, can induce both ferroelectric and ferromagnetic switching. While the coexistence of both the ferroic orders is encouraging from the view point of technological applications, the origin of ferromagnetism in this system remains elusive. Here, we synthesized such compositions and carried out a detailed structural analysis employing magnetic separation of the powder particles. We found that the origin of ferromagnetism lies in the spontaneous precipitation of a minor ( $\sim$ 1 wt %) barium hexaferrite (BaFe$_9$O$_{19}$) phase, often undetected in routine x-ray diffraction studies of powders sampled from the entire specimen. We also demonstrate that inspite of the insignificant fraction the ferrimagnetic phase, this self-grown composite exhibit noticeably enhanced ferroelectric switching in the presence of external magnetic field. We obtained a dc magnetoelectric coupling of $\sim$ 9 x 10$^{-8}$ s/m, a value which is comparable to what has been reported for layered ferroelectric/ferromagnetic laminates and bilayer thin film ferroelectric-ferromagnetic hetrostructures. Our study suggests that reasonably large magnetoelectric coupling is realizable in simple 0-3 ferroelectric-ferromagnetic bulk composites provided synthesis strategies are developed which induces spontaneous precipitation of the ferromagnetic phase in small volume fraction to ensure good insulating behaviour of the composite thus developed.

* ranjanrajeeb@gmail.com

# I. Introduction

Magnetoelectric multiferroic materials exhibiting coexistence of ferromagnetism and ferroelectricity are of great interest both from the scientific and technological viewpoints [1-3]. Most single phase multiferroics however exhibit antiferromagnetic ordering [4, 5]. As a consequence, even though these systems show intrinsic coupling between the antiferromagnetic order and polarization via the structural degrees of freedom [6, 7], they are not suitable for magnetoelectric devices as the applied magnetic field has no influence on the system. $BiFeO_3$ (BFO) is one of the most investigated room temperature multiferroic material [7-10]. The ferroelectricity in BFO arises due to off-centered displacement of the $Bi^{+3}$ along the [111] of the perovskite cell by virtue of the stereo chemical activity of its $6s^2$ lone pair electron [7]. The magnetic order, on the other hand, is associated with moments on the Fe-sublattice. In bulk BFO these moments form a long wavelength (~ 62 nm) cycloidal configuration superposed on a G-type antiferromagnetic order [11], and preclude a linear magnetoelectric coupling. Attempts have been made to suppress this long period modulation and induce a ferromagnetic correlation by size reduction [12], strain engineering in epitaxially grown films [8,9], and chemical modification [3, 6]. While, onset of ferromagnetism and electric-field induced changed in the magnetization has been demonstrated in epitaxial BFO films [9], the challenge persists in bulk specimens.

A popular strategy among experimentalists is to alloy BFO with other stable perovskites such as such $BaTiO_3$ [13-17], $PbTiO_3$ [18-21], rare-earth elements [22- 26], non-magnetic ions at the Fe-site [27], and look for improvements in the desired properties. Among others, (1-z)$BiFeO_3$-(z)$BaTiO_3$ (BF-BT) has attracted significant attention in the recent past [13-17]. The system exhibits complete solid solubility with rhombohedral structure in the composition range $0 \leq z < 0.33$, cubic-like for $0.33 \leq z \leq 0.92$, and tetragonal for $0.92 < z \leq 1$ [28]. Of particular interest in the context of magnetoelectric behaviour of this alloy system are the reports of ferromagnetism [6, 15, 29-31]. Although the coexistence of ferroelectricity and ferromagnetism in BF-BT is very interesting, exploitation of its magnetoelectric potential rests on our understanding of the origin of ferromagnetism in this system. While Park et al [15] argued that the ferromagnetism is intrinsic in nature and arise due to release of the latent magnetization locked in the long ranged cycloidal spin structure of the parent compound $BiFeO_3$ [15], other groups who reported ferromagnetism in this system [29-32] have not discussed/speculated its origin. Here, we examine the origin of ferromagnetism in the $BiFeO_3$-$BaTiO_3$ system in detail by carrying out a combined magnetization and structural study involving magnetic separation of the powder particles. We prove that the ferromagnetism is extrinsic in nature and owes its origin to spontaneous precipitation of a minor (~ 1 wt. %) barium hexaferrite phase, often undetected in routine x-ray diffraction of powder sampled from the whole specimen. We also demonstrate that inspite of the insignificant fraction of the ferromagnetic phase, the bulk specimen exhibit a very high dc magnetoelectric coupling $\alpha$ ~ 9 x $10^{-8}$ s/m, a value which is comparable to what has been reported for ferroelectric-ferromagnetic laminates [33] and thin films [34].

## II. Experimental

Following literature, we focused on synthesizing dense ceramics of BF-BT compositions exhibiting good ferroelectric and ferromagnetic properties using CuO and $MnO_2$ as additives [30-32]. For this study, we chose the composition $0.675BiFeO_3$-$0.325\ BaTiO_3$ and added $x$ wt % CuO + $y$ wt % $MnO_2$ as grain growth additives. The relative fractions of the two additives were varied as (i) $x$=0, $y$=0; (ii) $x$=0.1, $y$=0.15; (iii) $x$=0.2, $y$=0.30; and (iv) $x$=0.4, $y$=0.6. In the remaining of this paper, we will label these samples as Sample-1, Sample-2, Sample-3 and Sample-4, respectively. The ceramics were synthesized by conventional solid state reaction method using $BaCO_3$, $Fe_2O_3$, $Bi_2O_3$, $TiO_2$, CuO, and $MnO_2$ as raw materials of high purity grade (~99.9%, Alfa Aesar). The powders were mixed according to the stoichiometric formula and ball milled in acetone medium for 12 h to get a homogeneous mixture. The powders were calcined at 820 $^0$C for 2 h. The calcined powders were pressed into disks of 20 mm diameter and ~1 mm thickness using 2 wt % polyvinyl alcohol (PVA) as a binder. The pellets were sintered at 990 $^0$C for 2 h in closed alumina crucible with calcined powder of the same composition used as spacer. This sacrificial medium was used to reduce the evaporation of Bi during sintering. For electrical measurements, the opposite faces of the sintered pellets were painted with silver paste. X-ray powder diffraction study was carried out using a rotating anode diffractometer (Rigaku, SmartLab) equipped with a Johansson monochoromater in the incident beam to remove the Cu–Kα$_2$ radiation. Magnetization study was carried out using the vibrating sample magnetometer (VSM- LakeShore). Microstructural characterization was done with a scanning electron microscopy (ESEM Quanta 200). Polarization electric–field (*P-E*) loops were measured with a Precision Premier II loop tracer.

## III. Results

### A. *Structural, microstructural, ferroelectric and magnetic properties*

Figure 1 shows the XRD patterns of Sample-1, Sample-2, Sample-3, and Sample-4. In conformity with the earlier reports [30-32], the XRD patterns suggest formation of perovskite phase with no visible impurity peaks. The singlet nature of the Bragg profiles, confirms the structure to be cubic-like, in agreement with the reported structure for this composition [28]. The scanning electron micrograph (SEM) images revealed the average grain size to increase with increasing the additive content (from submicron grains - 0.6 *μm* for x=0, y=0 (Sample 1) to ~ 4 *μm* for *x*=0.4, *y*=0.6 (Sample 4), Fig. 2, confirming that these additives acts as grain growth agents. The enhancement in the ferroelectric switching with increasing additive contents, shown in Fig. 3(a), proves a direct correspondence between increased grain size and increased domain switching. Interestingly, and as was intended, the ferromagnetic behaviour of the specimen also improves dramatically with increasing the additive content with sample 4 showing a nice M-H hysteresis with a magnetization of ~ 0.45 emu/g at a 20 kOe, Fig. 3b. Sample 4 is therefore equivalent to the BF-BT specimens reported in the past by different groups exhibiting both ferroelectricity and ferromagnetism [30-32]. We chose this Sample for a detailed investigation.

### B. *Ferromagnetism: intrinsic or extrinsic?*

We examined the two scenarios: (i) ferromagnetism is intrinsic to the perovskite phase as proposed by Park et al [15], or (ii) ferromagnetism is extrinsic, and arise from another phase the volume fraction of which is so small that it is not revealed in XRD patterns of powder sampled from the entire specimen. To answer this puzzle, we adopted the strategy of magnetic separation of the powder particles. If the perovskite phase is ferromagnetic, then all the powder particles of the sample should be attracted towards a magnet, if placed in its proximity. On the other hand, if the ferromagnetism is due to another phase, then only those few particles would be attracted.

With this in view, we ground the sintered pellet of Sample 4 to fine powder and brought a magnet bar in its vicinity. We found that only a very small fraction of the ground powder got attached to the bar magnet. These powder particles were collected separately for structural analysis. Fig. 4a shows the XRD pattern of the particles collected on the bar magnet. The diffraction pattern clearly reveals additional Bragg peaks corresponding to a non-perovskite phase. Indexing of these non-perovskite peaks revealed the phase to be barium hexaferrite $BaFe_{12}O_{19}$. The high intensity peaks in this pattern correspond to the perovskite phase. This confirms that the particles attached to the bar magnet comprise of bonded grains of the perovskite and barium hexaferrite. For the sake of direct comparison, we also synthesized $BaFe_{12}O_{19}$, the XRD pattern of which is also shown in Fig. 4a. Figs. 4 (b) and (c) show the magnetic hysteresis data of the powders collected on the bar magnet, and $BaFe_{12}O_{19}$ respectively. We used the magnetization data of $BaFe_{12}O_{19}$ as a reference to estimate the weight fraction of the $BaFe_{12}O_{19}$ phase in Sample 4. At 20 kOe the magnetization of pure $BaFe_{12}O_{19}$ is 52 emu/gm (Fig. 4e), and that of the Sample 4 (before magnetic separation) is 0.45 emu/g (Fig. 3b). This suggests that the weight fraction of $BaFe_{12}O_{19}$ in Sample 4 is ~ 0.9 %. It is therefore not surprising that the Bragg peaks corresponding to the $BaFe_{12}O_{19}$ are not detectable in the XRD pattern shown in Fig. 1.

Having proved that the origin of ferromagnetism in our sample is extrinsic, in the next step we investigated the origin ferromagnetism in a lower $BaTiO_3$ content alloy, i.e. *0.9BiFeO₃-0.1BaTiO₃* reported in the past [6]. To be in conformity with the earlier studies, we did not add the grain growth additives CuO and $MnO_2$ during the synthesis of this composition. As anticipated for this composition [6], the splitting of the pseudocubic $\{111\}_{pc}$ Bragg profile into two (shown in the inset of Fig. 5a), confirms the rhombohedral structure. We also noticed a weak reflection (marked with * in Fig. 5a) corresponding to a non-ferromagnetic impurity phase $Bi_2Fe_4O_9$ often reported in the $BiFeO_3$ rich end of the BFO-based alloys. As shown in Fig. 5c, this specimen shows ferromagnetism with magnetization 0.12 emu/g at 20 kOe. As with Sample 4, we carried out magnetic separation of the powder particles of *0.9BiFeO₃-0.1BaTiO₃*. The XRD pattern of the magnetic rich particles, shown in Fig. 5b, confirms the additional Bragg peaks corresponding to the barium hexaferrite phase. In contrast to sample 4, however, the pellet of *0.9BiFeO₃-0.1BaTiO₃* did not show polarization-electric field hysteresis loop typical of a normal ferroelectric material due to significantly enhanced conductivity in conformity with the earlier studies [6].

### C. Magnetoelectric coupling

Having proved that the observed ferromagnetism in this system is extrinsic, i.e., they arise due to the precipitation of ~ 1 wt % barium hexaferrite grains in between ferroelectric grains, we investigated the coupling between the two phases and its response on macroscopic properties. We carried out polarization switching experiments under magnetic field, and also magnetization switching study before and after subjecting the specimen to strong external electric field. The experiments were carried out on sample 4 as it exhibited the best ferroelectric and ferromagnetic properties in our series. Fig. 6a shows the *P-E* loops of sample 4 at different values of dc magnetic field. The saturation polarization (*Ps*) increases from 20.83 $\mu C/cm^2$ at H = 0 kOe to 27.85 $\mu C/cm^2$ at 10 kOe i.e. by ~ 34 %, Fig. 6b. We also noted a slight decrease in the coercive field (*Ec*) from 22.5 kV/cm to 21.5 kV/cm when the field was increased to 10 kOe. The dc magnetoelectric response calculated using $\Delta Ps/\Delta H$ is ~ $8.81 \times 10^{-8}$ s/m. This value is comparable with the reported magnetoelectric coupling $\alpha$ of a $Pb(Zr,Ti)O_3$/terfenod-D laminate composite (~$10^{-8}$) [33] and bilayer epitaxial thin film of $BaTiO_3$/(La, Sr)$MnO_3$ [34] (~ 2 x $10^{-7}$ s/m).

We also carried out P-E measurements in the subcoercive field region. In this regime, the polarization can be approximately considered to be a linear function of field. Often P-E measurements in the subcoercive regime are performed to understand domain switching behaviour in the framework of Rayleigh analysis [35-38]. Rayleigh analysis assumes movement of domain walls in a medium comprising of pinning centers (defects) or varying strengths [35]. For small fields the domain wall motion is assumed to be reversible. The extent of the reversibility decreases as the coercive field is approached. The magnitude of the property (dielectric or piezoelectric) in the sub-coercive regime is proportional to the amplitude of the field [35].

$$\varepsilon_r = \varepsilon_{rev} + \alpha_\varepsilon * E_o \qquad (3)$$

where $E_o$ is the amplitude of the cyclic sub-coercive electric field. $\varepsilon_{rev}$ and $\alpha E_o$ represent the reversible and the irreversible contributions, respectively to the measured dielectric response. The dielectric coefficient is calculated from the P-E loops using the expression

$$\varepsilon_r = P_{p\text{-}p}/2E_o \qquad\qquad (4)$$

Where $P_{p\text{-}p}$ is peak-to-peak polarization measured for applied electric field amplitude $E_o$. For Rayleigh analysis P-E measurements are carried out at different electric field amplitudes to generate the ε - $E_0$ data. Here, our intention is not to carry out Rayleigh analysis on our specimen, but to examine the effect of magnetic field on the polarization of the specimen in the quasi-linear P-E regime. Accordingly, we fix the electric field amplitude of the measuring signal in the subcoercive region and varied the strength of the dc magnetic field applied parallel to the electric-field direction. The resultant plots are shown in Fig. 6c. The increase in polarization with increasing magnetic field is consistent with the trend observed even in the high field regime (Fig. 6a). The magnetic field dependence of the dielectric coefficient determined using equation 4 is shown in Fig. 6d.

## IV. Discussion

### A. *Sensitivity of the phase formed to synthesis conditions*

A perusal of literature suggests that there is no unanimity on the magnetic state in the BF-BT system. While ferromagnetic M-H loops has been reported by Park et al [15], Kim et al [17], and Wei et al [30]. No such loop is evident in the M-H data reported by Yang et al [14] and Buscagalia et al [39]. Park et al have claimed their specimens, prepared using molten salt synthesis method to be single phase and attributed the observed ferromagnetism to the release of the latent magnetization due to dissolution of the toroidal spin structure by finite size effect [15]. The authors reported maximum magnetization of 1.88 emu/g at 5 Tesla and a remanent magnetization of 0.75 emu/g in a composition z =0.2. For the same composition prepared by conventional ceramic synthesis method, Kim *et al.* also reported a ferromagnetic loop but with an order of magnitude lower remanent magnetization [17]. In contrast, similar compositions prepared by Yang *et al*., did not show ferromagnetic loop [14]. While Park et al have attributed the ferromagnetism in their specimen to the relatively small size and cube shaped particles, a perusal of the XRD data of (1-z)BiFeO₃-(z)BaTiO₃ presented in Fig. 2c of their paper reveals weak peaks on the higher 2θ side of the pseudocubic {110}$_{pc}$ doublet [15]. These peaks are not likely to be associated with the perovskite phase. The barium hexaferrite phase also shows a peak on the right side of the {110}$_{pc}$ in our specimen, Fig.4a. From this analogy we suspect that the additional peak in the XRD patterns of BF-BT specimens prepared by Park et al is also likely to be barium hexaferrite phase. That this peak can be seen in their pattern suggests that the volume fraction of the hexaferrite phase in their specimen is noticeably larger than in ours. We suspect that the ferromagnetism in their specimen is not likely to be intrinsic but due to the precipitation of the hexaferrite phase. As in our sample where in the ferromagnetic barium hexaferrite phase is precipitated by the CuO and MnO₂ additives, the used of salts in the molten salt synthesis method adopted by Park et al appears to assist the formation of the separate magnetic phase, not suspected by the authors. That this is also dependent on the concentration of the BaTiO₃ is evident from the fact that the hexaferrite phase is formed even without the additives in *0.9BiFeO₃-0.1BaTiO₃*. This is in conformity with the M-H data reported by Singh et al for the same composition [6]. The M-H data in ref. [6] however did not show saturation, presumably because the data was collected up to a maximum field of 6 kOe only. In view of our results, and the varied magnetic states reported by different research groups, it becomes evident that the precipitation of the barium hexaferrite phase in the BiFeO₃-BaTiO₃ system is very sensitive to the details of the synthesis conditions. A better understanding of the correlation between the synthesis conditions and the formation of the hexaferrite phase would enable better control of the volume fraction of the barium hexaferrite phase and the overall magnetoelectric behaviour of this material system.

### B. Significance of spontaneous precipitation of the magnetic phase

The significant increase in the dielectric response with increasing magnetic field (shown in Figs. 6c and 6d) suggests that domain wall mobility is significantly enhanced by the external magnetic field. In conventional Rayleigh analysis the dielectric coefficient increases with increasing amplitude of the measuring field (within the sub-coercive region). This is due to the increased ability of the domain walls to disentangle with the pinning centers. Since in our approach the electric field amplitude is fixed, the enhancement of the dielectric coefficient by the external magnetic field proves that magnetic field enhances ferroelectric-ferroelastic domain switching in the majority ferroelectric grains of our self-grown composite. The enhancement in the ferroelectric-ferroelastic switching with magnetic field is most likely to be associated with the transfer of the magnetostrictive strain from the barium hexaferrite grains to the surrounding ferroelectric grains. Our results suggest that the magnetostrictive stress increasingly depins the ferroelectric-ferroelastic domain walls in the ferroelectric grains. Our investigations have proved that inspite of extrinsic origin of the ferromagnetism in this system, electric field can significantly change the magnetization behaviour (the magnetic coercive field reduced by ~ 25 % after electric poling), and magnetic field could increase the polarization by ~ 33 % (when the magnetic field was increased to 10 kOe). These large changes suggest a very good coupling between the ferroelectric and the ferromagnetic grains. At this point we may caution that the elliptical loops observed in the P-E measurements in the subcoercive regime may as well be mimicked by leaky dielectrics. Such elliptical loops need not necessarily be associated with domain switching phenomena unless the materials exhibit saturation polarization, characteristic of a normal ferroelectric material, well above the coercive field.

While unlike single phase multiferroic systems, wherein the magnetoelectric coupling is fundamentally limited by the crystal and magnetic structure of the system, the magnetoelectric coupling in ferroelectric-ferromagnetic composites is determined by how efficiently the strain in one phase is transferred to the other. In this context, the quality of the contact between the ferroelectric and the ferromagnetic grains is an important factor. When the ferromagnetic phase spontaneously precipitates out in the piezoelectric matrix system it is likely to form a smooth and coherent interface with the neighboring piezoelectric grains, making stress/strain transfer very efficient. Moreover, in process involving spontaneous precipitation, the precipitated grains are likely to be homogeneously distributed over the entire specimen instead of getting clustered as is likely to be the case when the powders of the ferromagnetic phase is physically mixed with the piezoelectric powder to fabricate the composite. The fact that we could see a significant increase in the polarization even with ~ 1 wt % of hexaferrite phase confirms the efficient coupling between the ferroelectric and the ferromagnetic grains in our self-grown composite. We also performed magnetic hysteresis experiment before and after electric poling of Sample 4, and found the magnetic coercive field ($Hc$) to decrease substantially from 2700 Oe to 2000 Oe, i.e by

~ 26 %, Fig. 7. The value of the saturation magnetization was however not noticeably affected by electric poling. This experiment proves that electric poling has considerable influence on the magnetization behaviour of the embedded barium hexaferrite grains.

### C. Comparison with previous studies on in-situ grown ferroelectric-ferromagnetic composites

We may note that the idea of developing self-growing ferroelectric-ferromagnetic composites is not new. Run et al have reported *in-situ* grown $BaTiO_3$-$CoFe_2O_4$ ferroelectric-ferromagnetic eutectic composite [40]. They reported the ferromagnetic component to be ~ 40 volume percent and obtained dynamic magnetoelectric coupling coefficient ~ 50 mV/cm-Oe. Majumder and Bhattacharyya attempted to grow *in situ* $BaTiO_3$-$CoFe_2O_4$ composite but could achieve dynamic magnetoelectric coupling coefficient of ~ 5 mV/cm-Oe [41]. The low value is attributed to the formation of other unwanted phases during high temperature synthesis. A similar value was reported by Duong et al who developed core-shell microstructure of $BaTiO_3$-$CoFe_2O_4$ composite [42]. Since the oxide ferrimagnets are also semiconducting in nature, their presence in large quantity is likely to increase the leakage current of the composite specimen. It is therefore not surprising that most studies on bulk ferroelectric-ferromagnetic magnetoelectric composites do not show polarization-electric field loop characteristic of ferroelectric materials. Accurate estimation of quasi dc-magnetoelectric coupling as determined in our study is therefore not possible in such specimens. Although, at present, we do not understand the mechanism associated with the formation of barium hexaferrite phase in our system, the fact that additives such as $CuO$ and $MnO_2$ in our case are able to do so, is an encouraging development. Our results offer a hope that it might be possible to develop self-grown ferroelectric ferromagnetic composites in other systems as well by slight tweaking with the synthesis conditions, including additives which helps in controlled precipitation of a ferromagnetic phase. Such naturally grown ferroelectric-ferromagnetic composites are expected to perform better than the artificial fabricated composite due to reasons discussed above.

## V. Conclusions

In summary, we demonstrate that the ferromagnetic behaviour often reported in the $BiFeO_3$-$BaTiO_3$ multiferroic system arise due to precipitation of minor barium hexaferrite phase (~ 1 wt %) often unnoticed in conventional characterization of the specimens. The compositions exhibiting good ferroelectric and ferromagnetic switching are spontaneously formed ferroelectric –ferromagnetic 0-3 particulate composites. We found that inspite of the very insignificant fraction of the magnetic phase, this composite exhibit a significantly large change in polarization on application of magnetic field. The dc magnetoelectric coupling was found to be 9 x $10^{-8}$ s/m, a value comparable to that reported in bi-layer ferroelectric-ferromagnetic thin films. We argue that this extraordinary property arise due to the efficient transfer of the magnetostrictive strain from the embedded barium hexaferrite grains to the surrounding piezoelectric grains by virtue of the smooth interphase boundaries, and preservation of the insulating nature of composite due to small fraction of the semiconducting phase. Our study provides a guideline to design natural 0-3 ferroelectric ferromagnetic with remarkably improved magnetoelectric coupling.

**Acknowledgement:**

One of the authors (Amit Kumar) would like to thank UGC (New Delhi) for providing Dr D S Kothari Post Doc Fellowship for this work. R. Ranjan gratefully acknowledges the Nano Mission Programme of Department of Science and Technology, Govt. of India for financial support (Grant No. SR/NM/NS-1010/2015 (G)). Science and Engineering Research Board (SERB) of the Ministry of Science and Technology is also acknowledged for financial assistance (Grant no. EMR/2016/001457).

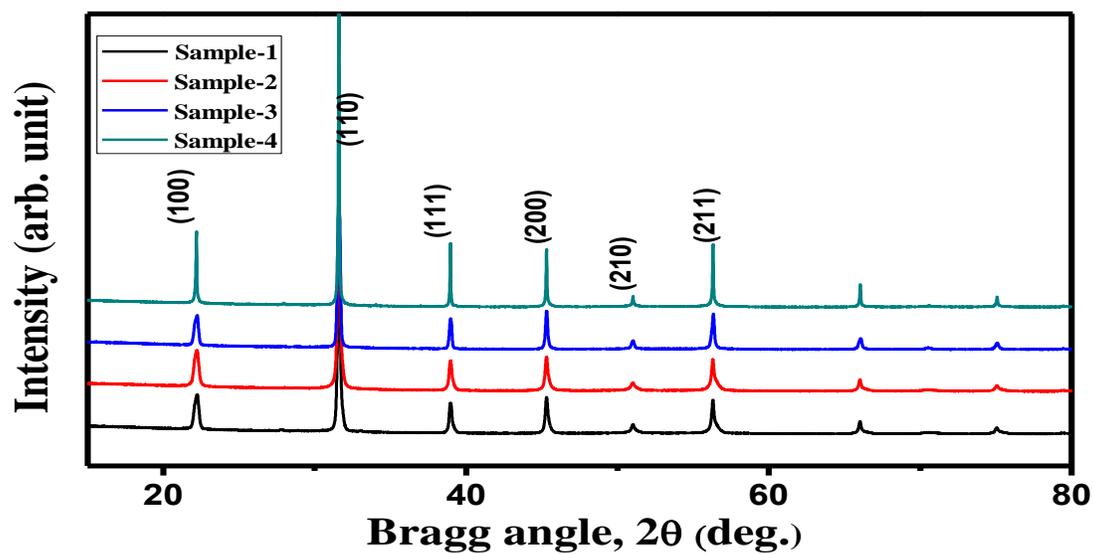

Fig. 1: X-ray powder diffraction patterns of 0.675BiFeO$_3$-0.325 BaTiO$_3$ synthesized with different CuO and MnO$_2$ additives. Sample 1 correspond to 0 wt % CuO and 0 wt % MnO$_2$, Sample 2 correspond to 0.1 wt % CuO and 0.15 wt. % of MnO$_2$, Sample 3 correspond to 0.2 wt % CuO and 0.3 wt % MnO$_2$ and sample correspond to 0.4 wt % CuO and 0.6 wt % MnO$_2$.

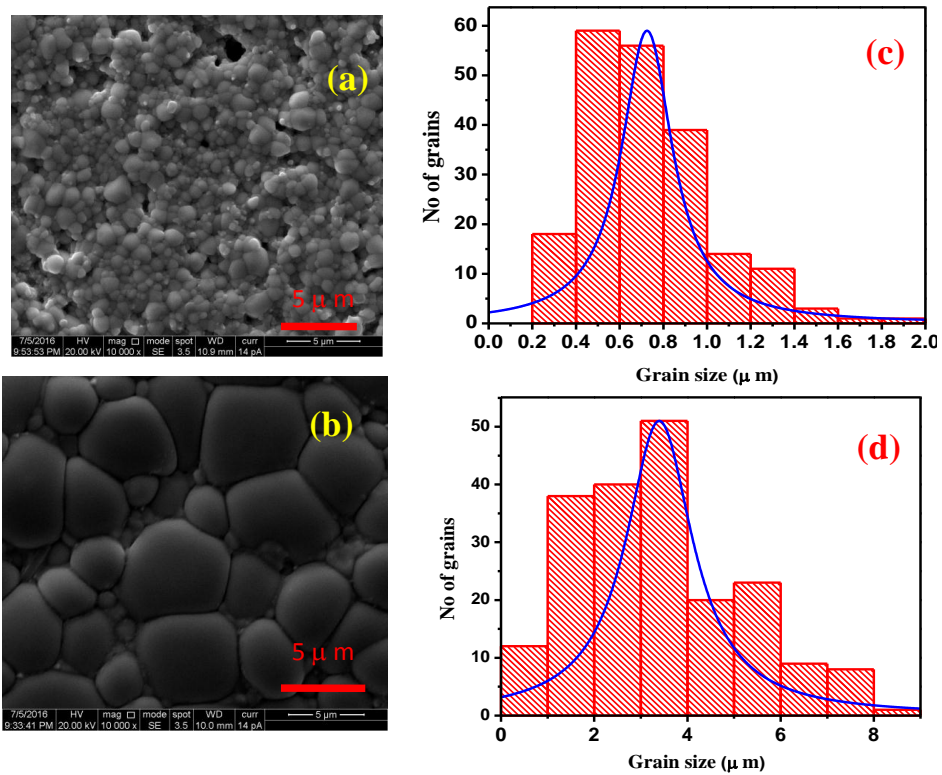

Figure 2: SEM images of (a) Sample-1 and (b) Sample-4. The histograms of the grain size distribution are shown on the right side of the respective SEM images

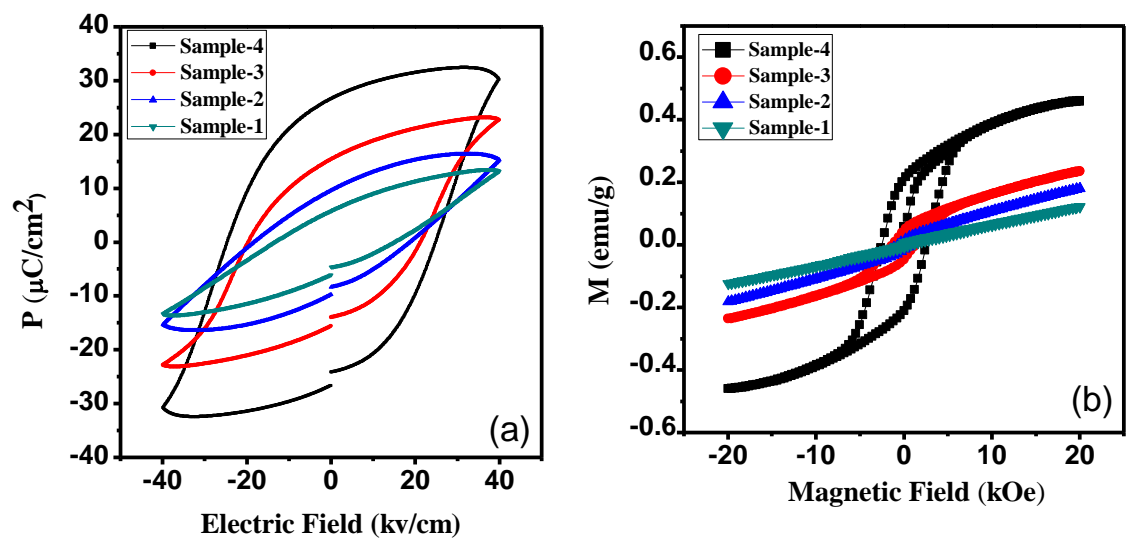

Figure 3 (a) Polarization-electric field hysteresis loops of sample-1 to sample-4. The magnetization –magnetic field hysteresis loops of the corresponding samples are shown in (b).

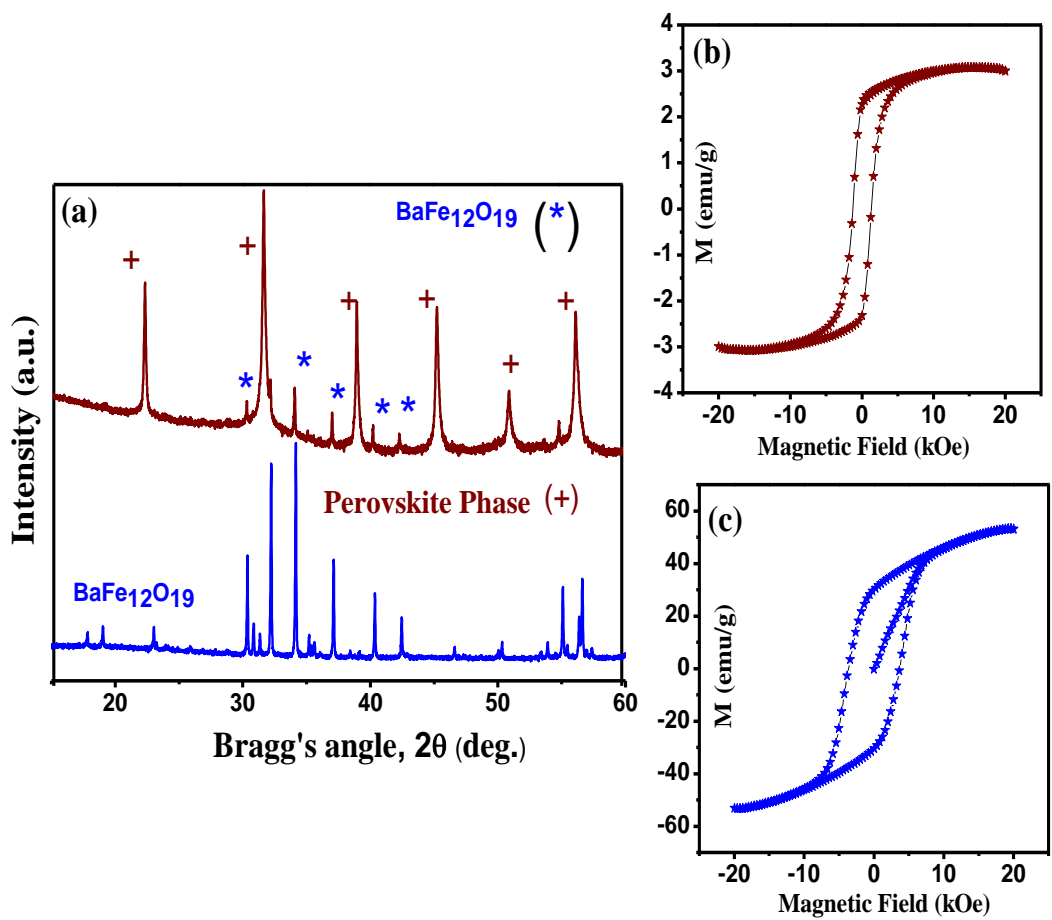

Figure 4 (a) shows the XRD patterns of the powder collected on the magnetic bar. For sake of convenience the Y axis is plotted on a logarithmic scale. The XRD pattern in the bottom panel of (a) is that of barium hexaferrite $BaFe_{12}O_{19}$. (b) shows the M-H loops of magnetically separated powder particles collected on the bar magnetic. (c) shows the M-H curve of the barium hexaferrite $BaFe_{12}O_{19}$.

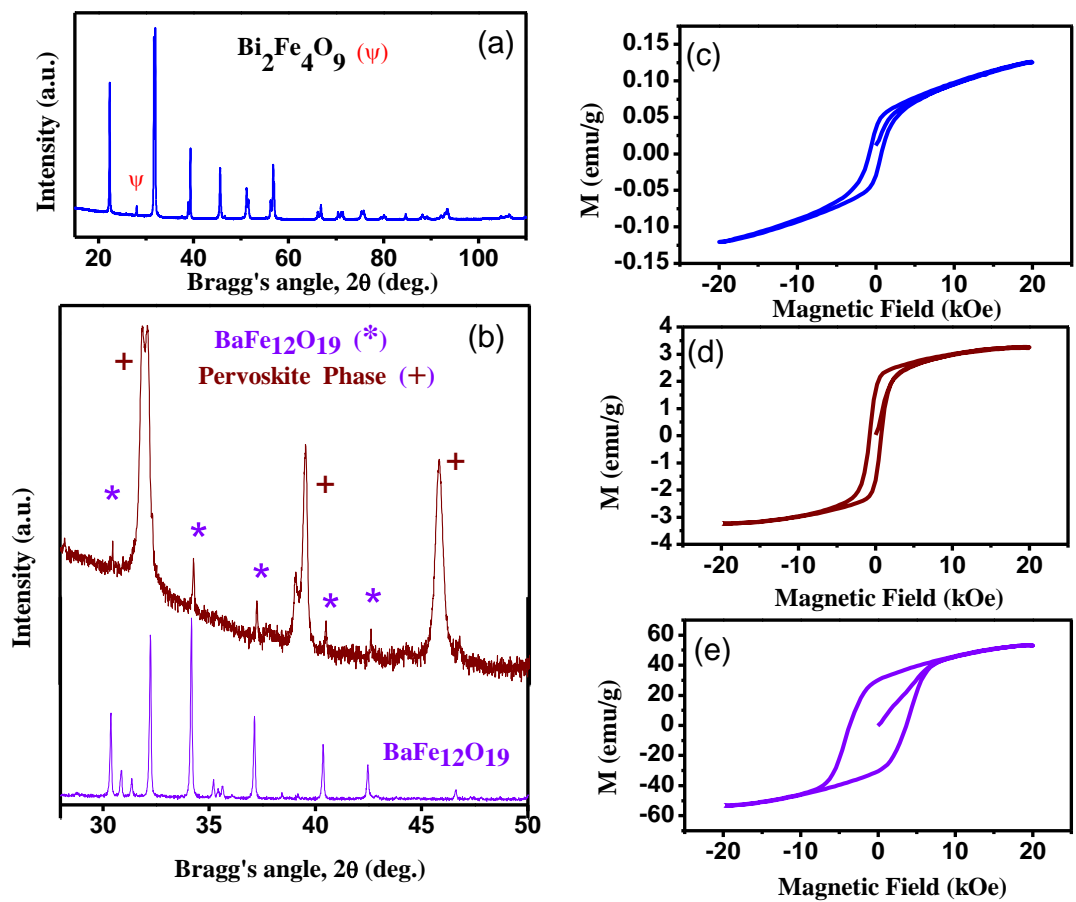

Figure 5, (a) shows the XRD patterns of 0.9BiFeO$_3$-0.1BaTiO$_3$. (b) shows the XRD patterns of the powder particles attached on the bar magnet (top), and barium hexaferrite (bottom). The Y-scale in (b) is shown on a logarithmic scale for easy visualization of the weak hexaferrite peaks. (c) –(e) shows the M-H loops of 0.9BiFeO$_3$-0.1BaTiO$_3$, magnetic rich phase, and BaFe$_{12}$O$_{19}$, respectively.

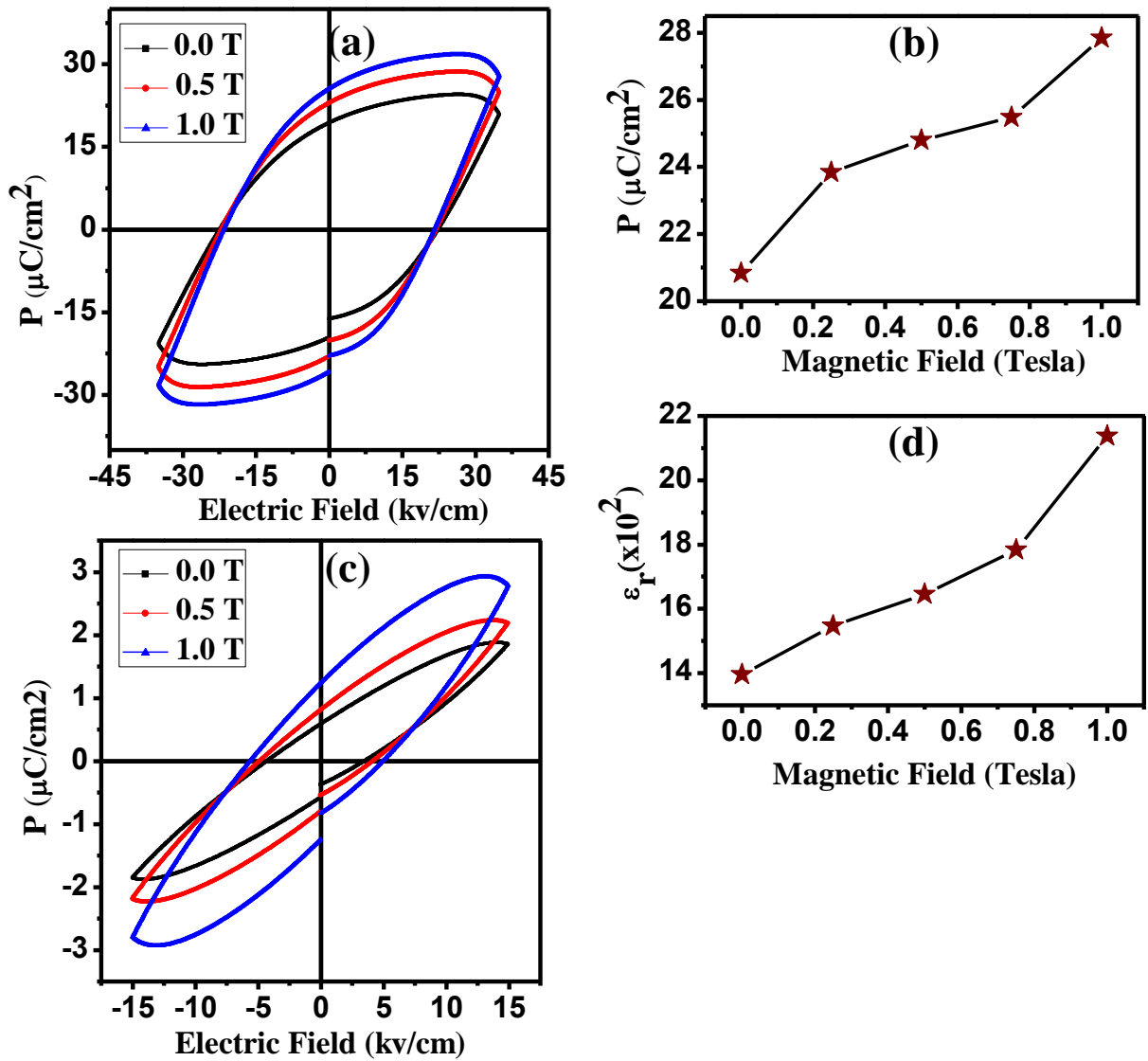

Figure 6 (a) shows the polarization – electric field hysteresis loops of sample-4 in the presence of magnetic field at room temperature. The maximum polarization as a function of magnetic field is shown in (b). (c) shows the (sub coercive electric field ) of sample-4 in the presence of magnetic field at room temperature (c) Variation of polarization in the presence of magnetic field, ( observed by part a), (d) variation of magneto electric coupling coefficient with magnetic field (calculated by using part b), (e) variation of dielectric constant with magnetic field (calculated by using part b)

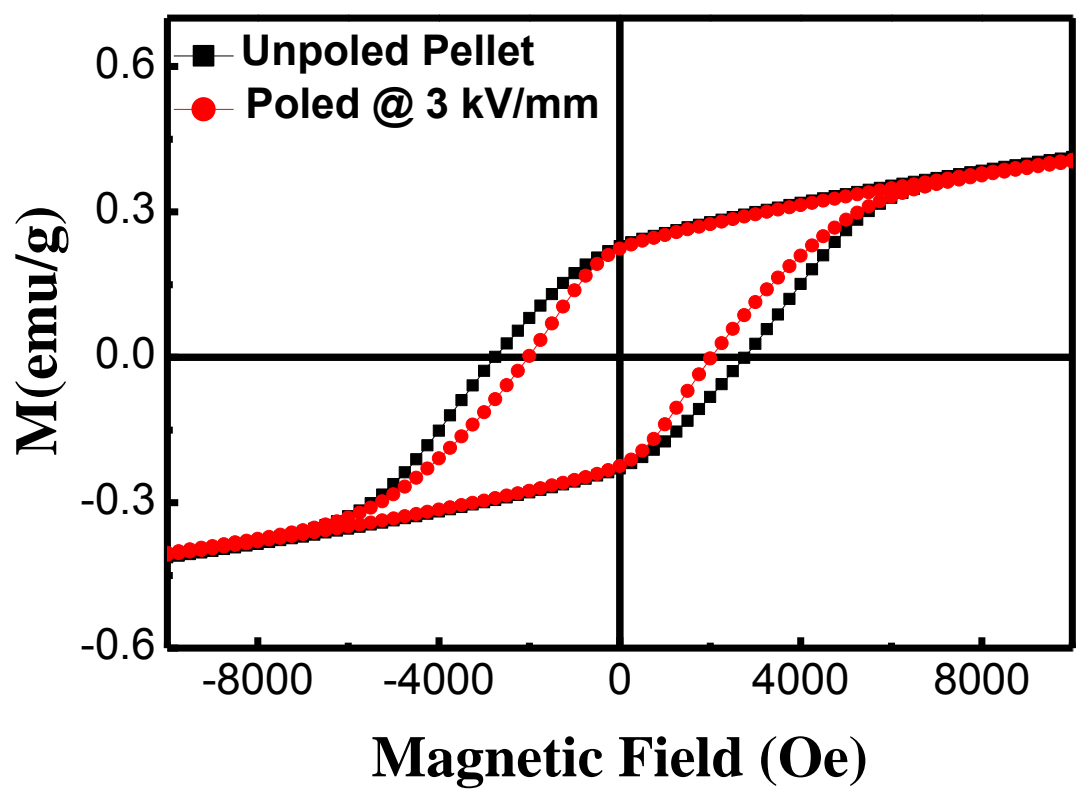

Fig. 7 Magnetization hysteresis curves of sample 4 before and after electric poling.